\documentclass[10pt,letterpaper]{article}
\usepackage[top=0.85in,footskip=0.75in]{geometry}

\usepackage{amsmath,amssymb}

\usepackage[utf8x]{inputenc}

\usepackage{textcomp,marvosym}

\usepackage{cite}

\usepackage{nameref,hyperref}

\usepackage{microtype}
\DisableLigatures[f]{encoding = *, family = * }
\usepackage{changepage}
\usepackage[table]{xcolor}

\usepackage{array}

\newcolumntype{+}{!{\vrule width 2pt}}

\usepackage[T1]{fontenc}
\usepackage{lmodern}
\usepackage{epsfig}
\usepackage{epstopdf}
\usepackage{graphicx}
\usepackage{multirow}

\newlength\savedwidth

\raggedright
\usepackage[aboveskip=1pt,labelfont=bf,labelsep=period,justification=raggedright,singlelinecheck=off]{caption}

\makeatletter
\renewcommand{\@biblabel}[1]{\quad#1.}
\makeatother

\usepackage{lastpage,fancyhdr,graphicx}
\usepackage{epstopdf}
\fancyheadoffset[L]{2.25in}
\begin{document}
	\vspace*{0.2in}
	
	\begin{flushleft}
		{\Large
			\textbf\newline{Think then act or act then think?} 
		}
		\newline
		\\
		Arkadiusz J\k{e}drzejewski\textsuperscript{1},
		Grzegorz Marcjasz\textsuperscript{2},
		Paul R. Nail\textsuperscript{3},
		Katarzyna Sznajd-Weron\textsuperscript{1,*}
		\\
		\bigskip
		\textbf{1} Faculty of Fundamental Problems of Technology, Wroc\l{}aw University of Science and Technology, Wroc\l{}aw, Poland
		\\
		\textbf{2} Faculty of Pure and Applied Mathematics, Wroc\l{}aw University of Science and Technology, Wroc\l{}aw, Poland
		\\
		\textbf{3} Faculty of Psychology and Counseling, University of Central Arkansas, USA
		\\
		\bigskip

		* katarzyna.weron@pwr.edu.pl
		
	\end{flushleft}
	\section*{Abstract}
	We introduce a new agent-based model of opinion dynamics in which binary opinions of each agent can be measured and described regarding both pre- and post-influence at both of two levels, public and private, vis-à-vis the influence source. The model combines ideas introduced within the $q$-voter model with noise, proposed by physicists, with the descriptive, four-dimensional model of social response, formulated by social psychologists. We investigate two versions of the same model that differ only by the updating order: an opinion on the public level is updated before an opinion on the private level or vice versa. We show how the results on the macroscopic scale depend on this order. The main finding of this paper is that both models produce the same outcome if one looks only at such a macroscopic variable as the total number of the individuals with positive opinions. However, if also the level of internal harmony (viz., dissonance) is measured, then significant, qualitative differences are seen between these two versions of the model. All results were obtained simultaneously within Monte Carlo simulations and analytical calculations. We discuss the importance of our studies and findings from three points of view: the theory of phase transitions, agent-based modeling of social systems, and social psychology.

	
	\section*{Introduction}
	There is a common belief, verbalized by Ralph Waldo Emerson in his Essays from 1841, that ``The ancestor of every action is a thought''. On the other hand, as noted by social psychologist David Myers \cite{Mye:10}: ``If social psychology has taught us anything during the last 25 years, it is that we are likely not only to think ourselves into a way of acting but also to act ourselves into a way of thinking.''. Indeed, it has been shown in many social experiments that attitudes are frequently poor predictors of behaviors, and it is often that behaviors determine attitudes \cite{Mye:10}. 
	This startling conclusion inspired us to build a microscopic model with private and public opinions that combines the idea introduced within the $q$-voter model with independence \cite{Nyc:Szn:Cis:12}, an agent-based model (ABM) of opinion formation, with the descriptive, four-dimensional model of social response \cite{Nai:Mac:Lev:00}.
	
	According to the latter model, the position or opinion of each individual can be measured and described regarding both pre- and post-influence at both of two levels, public and private, vis-à-vis the influence source. The introduction of these variables  allowed for the distinction between the majority of the influence types that are generally recognized as theoretically significant and conceptually distinct. For example, they provide for the distinction between two widely recognized types of conformity: conversion (conformity at both public and private levels) and compliance (public conformity without private acceptance). Moreover, it allows for the description of a particularly interesting type of social response, so-called disinhibitory contagion, which is similar to conformity in that, in both cases, an individual changes his or her public behavior toward that of an external influence source. However, the reason that disinhibitory contagion is very different than conformity is that the former begins with an influencee's internal intra-psychic conflict before even being exposed to the influence source. 
	
	To understand the essence of disinhibitory contagion, imagine the following situation. You and your date are at a dance, but no one is dancing. You and your partner are inhibited from dancing because you would be embarrassed to be the only couple on the floor. This is the ``force'' that prevents you from doing what you would like to do. At this point, however, one couple starts dancing (viz., the initiators or triggers), then another and another. You and your date join in, and the dance floor quickly fills. Note that the influence of the initiators is contagious, as the influence spreads, and the situation progresses from your initial internal conflict over dancing $\rightarrow$ through influence $\rightarrow$ to your ultimate internal harmony, happy that you are dancing. In this case, it is only your and the other agents' public behavior that changes, not private opinions. Another example of disinhibitory contagion may be the situation in which Poland initiated the democratic reforms in defiance of the historic domination of the Soviet Union that swept across Eastern Europe in 1989. 
	
	In the field of sociophysics, various models of binary opinion dynamics have been proposed \cite{Cas:For:Lor:09,Gal:12,Sen:Cha:13,Sir:Lor:Ser:17,Lew:Now:Lat:92,Hol:Kac:Sch:01,Gal:90,Wat:02,Kra:Red:03,Mel:Mob:Zia:17,Fer:etal:14}, but only a few of them take into account the distinction between 
	private (PRIV) opinions (or attitudes) and public (PUB) behaviors. Probably such a two-level description was introduced for the first time within the Continuous Opinions and Discrete Actions (CODA) model \cite{Mar:08}. The model was later developed and modified in various directions \cite{Mar:Gal:13,Mar:14}. However, in all its versions discrete public choices are resulting from the private continuous opinions in such a way that the inner continuous opinion is used to measure how certain each agent is about its decision on which of two options it prefers. This means that the CODA model describes only rational individuals that act in accordance with their own convictions, which as we know from social psychology, is not always true. 
	Another model that differentiates between public and private opinions has been proposed very recently under the name concealed voter model (CVM) \cite{Gas:Obo:Gul:18}. It is built upon the linear voter model \cite{Cli:Sud:73} by adding a concealed, private, layer of opinions to the public layer present in the original model. In the concealed voter model, both opinions are binary, and in the case of disparity between them, one opinion can be embraced by the other with certain rates. On the complete graph of a finite size, agents within CVM must reach a consensus, although the consensus time is longer than for the basic voter model \cite{Cli:Sud:73}.
	
	The model introduced here, on the other hand, covers more possibilities of behaviors allowing for not only various ``rational'' but also  ``irrational'' ones in comparison to the CODA model, and it also takes into account the role of the group influence in the opinion making process in contrast to the concealed voter model. Moreover, it has strong foundations in social psychology. In fact,  the model is directly derived from one of the most sophisticated descriptive models of social response. However, introducing yet another agent-based model of opinion dynamics is not the aim of this paper. Rather, herein we ask two questions, and we hope that the answers will be important not only for social agent-based modeling but also for social psychology. 
	
	Keeping in mind that a basic goal of theorists and researchers in the field of descriptive models of social response is to identify the minimum number of variables that are needed to adequately distinguish between as many of the important types
	of social influence as possible, we can ask if two levels (private and public) of description are really necessary. Of course, the question formulated in this way is very imprecise. However, we can ask this question rigorously within a specific ABM, here the $q$-voter model, in the following way: Will the $q$-voter model with private and public opinions give qualitatively different results than the original $q$-voter model with only public opinion? 
	
	Another question that can be asked has been already posed in the title of this paper and is closely related to the first sentences of the Introduction. What comes first -- thinking or acting? Social psychologists try to answer the following  questions \cite{Mye:10}: ``How do our inner attitudes relate to our external behavior?'', ``Why does our behavior affect our attitudes?'', etc. Here, we will ask another question inspired by the above but more closely related to the agent-based modeling procedure. While building an ABM, it is necessary not only to define precisely agents, as well as a network in which they exist and interact with each other. It is also necessary to give a precise algorithm that describes how the system evolves over time. This means that we have to decide, for example, about the order of updating. Usually, when building a social ABM, we rely on a social theory and experiments. In this case, however, social psychology tells us that there is no simple answer to the question: What should be updated first - a public or a private opinion? Therefore, we investigate herein two versions of the model. One in which we update (a) the private opinion first and then the public one, and the second version in which the opposite updating scheme occurs (b) the public opinion is updated first and then the private one. Because often a public opinion can be identified with the way of acting and a private opinion with the way of thinking, we call these two variants of the model: (a) think then act (TA) versus (b) act then think (AT). Now, we are ready to formalize the precise question, which is the focal point of this work: To what extent will the TA and AT models give different results at the macroscopic level?
	
	\section*{Model}
	
	The ABM we introduce here is based on a descriptive model of social influence, the so-called four-dimensional (4D) model  \cite{Nai:Mac:Lev:00}. Its roots are from the Willis  $+/-$ scheme for symbolizing responses to social influence \cite{Wil:63}. Within the scheme, we describe the response to social influence in a single set of interactions (a single social influence trial) by a series of 3 signs: the first sign ($+$ or $-$) represents the target’s initial  position (the pre-influence opinion), the second ($+$ or $-$) the position advocated by the potential influence source, and the third ($+$ or $-$) the target’s response to the source (the post-influence opinion). This means that in total we have $2^3=8$ combinations, which can be grouped into four basic response patterns:
	\begin{eqnarray}
	\mbox{C}:& + -- \hspace{0.2cm} \mbox{or} \hspace{0.2cm} - ++ ,\nonumber\\
	\mbox{I}:& + - + \hspace{0.2cm} \mbox{or} \hspace{0.2cm} - + - ,\nonumber\\
	\mbox{A}:& + + - \hspace{0.2cm} \mbox{or} \hspace{0.2cm} - - + ,\nonumber\\
	\mbox{U}:& + + + \hspace{0.2cm} \mbox{or} \hspace{0.2cm} - - -, 
	\end{eqnarray}
	where $C$, $I$, $A$, and $U$ stand for conformity, independence, anticonformity, and uniformity respectively. 
	Now, if we assume that these pre- and post-influence opinions can occur independently at both private and public levels, as in the
	4D model \cite{Nai:Mac:Lev:00}, we obtain $32$ combinations.  However, the consideration of symmetry allows  these combinations to be reduced to 16 different response patterns of the four-dimensional model\cite{Nai:Mac:Lev:00} (see Table \ref{tab:model_4d}).
	
	\begin{table}
		\begin{center}	
			\begin{tabular}{llllll}
				\multirow{2}{*}{\#1} & \multirow{2}{*}{Congruence} & PUB & $+++$ & \multirow{2}{*}{or} & $---$\\
				&& PRIV & $+++$ & & $---$\\
				\multirow{2}{*}{\#2} & \multirow{2}{*}{Paradoxical Compliance} & PUB & $+++$ & \multirow{2}{*}{or} & $---$\\
				& & PRIV & $++-$ & & $--+$\\	
				\multirow{2}{*}{\#3} & \multirow{2}{*}{Anticompliance} & PUB & $++-$ & \multirow{2}{*}{or} & $--+$\\
				&& PRIV & $+++$ & & $---$\\	
				\multirow{2}{*}{\#4} & \multirow{2}{*}{Anticonversion} & PUB & $++-$ & \multirow{2}{*}{or} & $--+$\\
				&& PRIV & $++-$ & & $--+$\\	    
				\multirow{2}{*}{\#5} & \multirow{2}{*}{Compliance/Conversion} & PUB & $+++$ & \multirow{2}{*}{or} & $---$\\
				&& PRIV & $-++$ & & $+--$\\	
				\multirow{2}{*}{\#6} & \multirow{2}{*}{Continued Compliance} & PUB & $+++$ & \multirow{2}{*}{or} & $---$\\
				&& PRIV & $-+-$ & & $+-+$\\  
				\multirow{2}{*}{\#7} & \multirow{2}{*}{Reversed Anticompliance} & PUB & $++-$ & \multirow{2}{*}{or} & $--+$\\
				&& PRIV & $-++$ & & $+--$\\ 
				\multirow{2}{*}{\#8} & \multirow{2}{*}{Disinhibitory Anticonversion} & PUB & $++-$ & \multirow{2}{*}{or} & $--+$\\
				&& PRIV &$-+-$ & & $+-+$\\  
				\multirow{2}{*}{\#9} & \multirow{2}{*}{Disinhibitory Contagion} & PUB & $-++$ & \multirow{2}{*}{or} & $+--$\\
				&& PRIV &$+++$ & & $---$\\   
				\multirow{2}{*}{\#10} & \multirow{2}{*}{Reversed Compliance} & PUB & $-++$ & \multirow{2}{*}{or} & $+--$\\
				&& PRIV &$++-$ & & $--+$\\   
				\multirow{2}{*}{\#11} & \multirow{2}{*}{Inhibitory Independence} & PUB & $-+-$ & \multirow{2}{*}{or} & $+-+$\\
				&& PRIV &$+++$ & & $---$\\
				\multirow{2}{*}{\#12} & \multirow{2}{*}{Anticontagion} & PUB & $-+-$ & \multirow{2}{*}{or} & $+-+$\\
				&& PRIV &$++-$ & & $--+$\\ 
				\multirow{2}{*}{\#13} & \multirow{2}{*}{Conversion} & PUB & $-++$ & \multirow{2}{*}{or} & $+--$\\
				&& PRIV &$-++$ & & $+--$\\    
				\multirow{2}{*}{\#14} & \multirow{2}{*}{Compliance} & PUB & $-++$ & \multirow{2}{*}{or} & $+--$\\
				&& PRIV &$-+-$ & & $+-+$\\  
				\multirow{2}{*}{\#15} & \multirow{2}{*}{Paradoxical Anticompliance} & PUB & $-+-$ & \multirow{2}{*}{or} & $+-+$\\
				&& PRIV &$-++$ & & $+--$\\      
				\multirow{2}{*}{\#16} & \multirow{2}{*}{Independence} & PUB & $-+-$ & \multirow{2}{*}{or} & $+-+$\\
				&& PRIV &$-+-$ & & $+-+$
			\end{tabular}
			\caption{The 16 possible response patterns introduced within the four-dimensional model of social response \cite{Nai:Mac:Lev:00}.}
			\label{tab:model_4d}
		\end{center}
	\end{table}
	
	Let us explain here two issues because we realize that they may not be clear to some readers. First of all, one should note  that in Table \ref{tab:model_4d} we use convention in which the private opinion of the source of influence is the same as its opinion on the public level. Of course, in general this is not necessarily true that the source of influence has the same opinion on both levels. However, it does not matter what opinion the source has at the private level because this level is not seen by the target. In fact, we could use another convention to describe the private level using, e.g., question mark in the place of '$+$' or '$-$' for the private opinion of the source of influence. Within such a convention, e.g., disinhibitory contagion would be defined as (PUB $+--$, PRIV $-?-$), instead of (PUB $+--$, PRIV $---$). For the order, we decided to stay with the original notation, introduced in \cite{Nai:Mac:Lev:00}. Second issue to be explained concerns the name of the model: ``Why is it called four-dimensional?''.  Notice that according to this model 4 variables (dimensions) have to be measured in order to identify the type of social response of the individual: pre-influence public opinion, pre-influence private opinion,  post-influence public opinion, and  post-influence private opinion. Measuring these 4 variables we can identify 16 different response patterns, as described above.
	
	Of course, the probabilities of these 16 responses and factors influencing a given type of response are different. Let us go back to the example with dancing given in the Introduction and denote dancing by $+$ and not dancing by $-$ . In this example, the influence of the initiators is contagious, as the influence spreads, and the situation progresses from internal conflict $\rightarrow$ influence $\rightarrow$ internal harmony. Note that it is the agents’ public behavior that changes, not their private opinion. This type of influence is dubbed disinhibitory contagion (PUB $-++$, PRIV $+++$) in the four-dimensional model \cite{Nai:Mac:Lev:00}. Because this type of influence is contagious the source of influence can consist of even only one person to effectively change the public opinion of the target, which is the same as assumed in the linear voter model  \cite{Fer:etal:14}. However, for other types of social response  it may not be true, i.e., more people are needed to influence the target. 
	
	To clarify this statement, let us once again use an example with dancing, but this time regarding conformity, specifically the four-dimensional model’s $\#14$ compliance conformity: You are at a dance even though you  do not like to dance and are not very good at it. At some point, you are pressured to dance by peers and end up being herded onto the dance floor. Using Willis's symbols, compliance conformity is represented by PUB $-++$ and PRIV $-+-$. Here, the situation progresses from initial internal harmony  $\rightarrow$ through influence  $\rightarrow$ to ultimate internal conflict. This type of conformity was the subject of research in the famous experiment with lines conducted for the first time by Asch and then repeated by many others \cite{Asch:56,Bon:05}. In such a case, the influence only from one person  may not be sufficient, and unanimity of opinions in the group of influence is critical. This observation was implemented in the original $q$-voter model \cite{Cas:Mun:Pas:09} and later on in its modified version with noise \cite{Nyc:Szn:Cis:12}. 
	
	In the original $q$-voter model, a target agent responds to the source of influence that consists of $q$ agents randomly chosen from the target's neighbors only if all $q$ agents have the same opinion \cite{Cas:Mun:Pas:09}. Therefore, the original formulation of the $q$-voter dynamics captures only one type of social response, that is, conformity, i.e., changing to the position of the source of influence. Subsequently, two other types of response, two types of nonconformity, were introduced: anticonformity and independence. The model with anticonformity will not be considered in this paper, and therefore, we will not discuss it here.  We are aware that by excluding anticonformity from the model as a driving interaction we may loose some generality. However, until now the $q$-voter model with one level of opinion (public) has been investigated with only one type of nonconformity: either anticonformity or independence, but not with both simultaneously.  Moreover, the $q$-voter model with independence has shown much richer and interesting behavior than the $q$-voter model with anticonformity \cite{Nyc:Szn:Cis:12,Nyc:Szn:13}. Therefore, to check if the second level of opinion introduces any new quality on the macroscopic level, i.e., to compare it with the one-level, public-only version, we limit our study to only one type of nonconformity, namely independence.
	
	Independence in social psychology means an absence of influence, and indeed, it means the same in the $q$-voter model, i.e., a voter does not change under the influence of others \cite{Nyc:Szn:Cis:12}. 
	However, independence does not mean that the agent must stay with its original opinion.  We have assumed that the agent can randomly change its opinion to the opposite one, independently of the neighborhood. In \cite{Nyc:Szn:Cis:12} the probability of such a change was $f=1/2$, which is quite high and maybe not very realistic. However, as shown analytically and numerically, the parameter $f$ only rescaled the results, and therefore, we can choose an arbitrary value as long as $f>0$ \cite{Szn:Tab:Tim:11}. The introduction of a noise in the form of this independent behavior prevents the complete consensus, which makes the model more realistic. Moreover, the $q$-voter model with noise occurred to be very interesting from the physical point of view, displaying tricriticality, that is to say, a change in phase transition type, and surprising behavior on multi-layer networks \cite{Nyc:Szn:Cis:12,Chm:Szn:15}. For all of the above reasons, we have chosen the $q$-voter model with independence as the starting point for a more complex model based on the 4D model.
	
	Having in mind the role of internal harmony/disharmony, we propose the following model. We consider a network of $N$ nodes, each occupied by exactly one agent characterized by two variables: opinion described both on a public level ($S_i=\pm 1$) and on a private level ($\sigma_i=\pm 1$). In short, we will call them public and private opinions. Opinions $S_i$ and $\sigma_i$ of $i$-th agent are dynamic variables, which means that both of them consecutively can change in time under social influence or noise in the form of independence. We analyze and compare both updating order possibilities. In one, we first update the public opinion, and then the private one (AT model); in the other, the updating is the opposite (TA model).

	The influence of independence and conformity relies on the type of opinion level they affect, and it is different for public or private one. Independence substitutes the value of the public opinion for the value of the private one, and it changes the private opinion to the opposite one with probability $1/2$, which corresponds to the noise introduced in the original $q$-voter model \cite{Nyc:Szn:Cis:12}. So on the public level, independence prompts us to express our private opinion publicly while on the private level, it makes us reconsider changing our private opinion independently of the neighbors. On the other hand, in the case of conformity, first we choose randomly $q$ neighbors of the agent who is exposed to the influence. The group is called $q$-panel, and it tries to exert social pressure on the chosen agent. Now, when we try to change the public opinion, and the chosen agent is in the internal conflict, $S_i\neq\sigma_i$, then we do not require unanimity of the agent's peers, and we substitute the value of $S_i$ for the value of $\sigma_i$ if at least one of all $q$ neighbors supports publicly our private opinion, i.e., has the same public opinion as the private opinion of the agent $i$. 
	Otherwise, when the agent is in the internal harmony, $S_i = \sigma_i$, we assume that it is harder to convince the target, and the social influence must be stronger in order to change the agent's public opinion since then its state goes over to the internal conflict. That is why the group of influence is successful and exerts social pressure only if it is unanimous, i.e., all members of the panel have the same public opinion. Then, the considered agent changes its public opinion to the public opinion shared by the group members.
	We deal with the similar situation in the case of conformity on the private level, but now we do not differentiate between the situations with internal conflict and harmony. In both scenarios, we require a unanimous group of influence to change the private opinion of the chosen agent since we assume that it is harder to change our believes than behaviors. So, the target agent changes its private opinion to the public opinion shared only by the members of the unanimous group.
	
	Furthermore, the level of independence $p$, which is assumed to be a static parameter of the model,  determines the probability that an agent acts independently. With complementary probability $1-p$, conformity occurs. Note that within this approach, each agent can be  either conformist or independent in different time steps, so the way of agents' acting may change during simulation. This corresponds to the so-called situation-oriented approach in which the social response of agents is determined by situational factors and is not connected with their personal traits \cite{Szn:Szw:Wer:14, Jed:Szn:17, Kru:Szw:Wer:17}. Another, competitive approach called person-oriented, which will not be considered here, assumes that it is our personality that dominates over the situation. Therefore, at the beginning of simulations, each agent is determined whether it is conformist or independent with the same probabilities as in the case of the situation approach, but now the agents' way of acting does not change in time. At first glance, these approaches may seem to give the same results since in both the probability of choosing an agent from the system that will act independently is the same and equals to $p$. Nevertheless, they may give, in fact, qualitatively different outcomes depending on the type of social response that is considered, as it is in the case of the $q$-voter model with independence but not with anticonformity \cite{Jed:Szn:17}.
	From this point of view, it would be interesting to analyze the model introduced here in the person-oriented approach in a future study, as well.

	In order to better comprehend the model, below we present the precise algorithm for the AT model in which we first update the public opinion:
	\begin{enumerate}
		\item At a given time step $t$, choose one voter at random, located at
		site $i$.
		\item Update the public opinion $S_i$:
		\begin{enumerate}
			\item With probability $p$, the agent acts independently, i.e., replace public opinion by the private one $S_i \rightarrow \sigma_i$.
			\item With complementary probability $1-p$, the agent is influenced by the public opinions of its $q$ randomly picked neighbors without repetition. If voter's public opinion is different than the private one, $S_i \ne \sigma_i$, then it changes its public opinion to the private $S_i \rightarrow \sigma_i$ if at least one of $q$ neighbors has the same public opinion as the private opinion of the agent $i$. Otherwise, i.e., if voter's public opinion is already the same as the private one, $S_i= \sigma_i$, the influence occurs only if all $q$ neighbors have the same public opinion. Then the agent $i$ adopts their opinion, i.e., $S_i \rightarrow S_j$, where $S_j$ is the public opinion of one among $q$ neighbors.
		\end{enumerate} 
		\item Update the private opinion $\sigma_i$:
		\begin{enumerate}
			\item With probability $p$, the agent acts independently, i.e., it randomly changes its private opinion $\sigma_i \rightarrow -\sigma_i$ with probability $1/2$.
			\item With complementary probability $1-p$, the agent is influenced by its $q$ randomly picked neighbors without repetition, but only by their public opinions. If all $q$ neighbors have the same public opinion then the agent $i$ adopts their opinion, i.e.,  $\sigma_i \rightarrow S_j$.
		\end{enumerate}
		\item Time is updated $t\rightarrow t+1/N$.
	\end{enumerate}
	The TA model differs from AT only in the way we switch the order of points 2 and 3 in the above algorithm, i.e., first we update the private and then the public opinion.
	Note that both versions of the model use the current, most recent value of the private opinion to update the state of the opinion at the public level in the case of independent behavior of an agent, look at the second step in the above algorithm. But since the public opinion is updated first in AT model, thus if time is included explicitly in the notation then we should write $S_i(t+1/N)=\sigma_i(t)$ whereas in TA model, we have $S_i(t+1/N)=\sigma_i(t+1/N)$ because the private opinion has been already updated.
	
	Let us now define quantities that we are interested in. The number of individuals with the positive opinion at  the public level is denoted by $N_S(t)$. Analogously, $N_{\sigma}(t)$ denotes the number of individuals with the positive opinion at the private level. Using Monte Carlo simulations, we measure 3 aggregated quantities:
	\begin{itemize}
		\item the fraction of individuals with the positive public opinion:
		\begin{equation}
		c_S(t)=\frac{N_S(t)}{N}=\frac{1}{2N}\sum_{i=1}^N(1+S_i(t)),
		\label{eq:def:pub}
		\end{equation}
		\item the fraction of individuals with the positive private opinion:
		\begin{equation}
		c_{\sigma}(t)=\frac{N_{\sigma}(t)}{N}=\frac{1}{2N}\sum_{i=1}^N(1+\sigma_i(t)),
		\label{eq:def:priv}
		\end{equation}
		\item the level of dissonance defined as a fraction of individuals that have different public and private opinions:
		\begin{equation}
		d(t)=\frac{1}{2N}\sum_{i=1}^N (1-\sigma_i(t)S_i(t)).
		\label{eq:def:dis}
		\end{equation}
	\end{itemize}
	
	We assume that initially nobody is in the internal disharmony, i.e., $S_i(0)=\sigma_i(0)$ for $i=1,2,\ldots,N$, so at the beginning of all simulations the level of dissonance equals to zero $d(0)=0$. 
	Furthermore, most of the results presented in this paper have been obtained from ordered initial conditions $S_i(0)=\sigma_i(0)=1$. However, other initial conditions have been also investigated, and some of results will be also presented here.
	The reason why we checked different initial conditions is that they might have an impact on the model behavior and its final state. For instance, in the original $q$-voter model with independence, there is a continuous phase transition for $q<6$. This means that there is no hysteresis, so the final state of a system is independent of initial conditions \cite{Nyc:Szn:Cis:12}. However, for $q \ge 6$, there is a discontinuous phase transition. The hysteresis appears, and hence, initial conditions start to play important role in the ultimate state reached by the system.

	\section*{Methods and results}
	One of the most popular methods for analyzing microscopic models of interacting many-particle systems -- or in another language ABMs -- is Monte Carlo computer simulation. Such simulations allow one to obtain an estimation of the expected values of measured quantities based on averaging over samples (ensemble average) or over time. In our research, we used the ensemble average. Although an arbitrary graph can be used as the underlying network for our model, and Monte Carlo simulations will deal with that, we consider only large ($10^5-10^6$ nodes) complete graphs since then we are able to carry out analytical calculations in the spirit of the mean-field approximation (MFA). Moreover, these calculations become strict in the limit $N\rightarrow\infty$ for such networks. This allow us to compare our analytical predictions with the simulations. A comprehensive description of the mean-field approximation together with its application to the less complicated, original $q$-voter model with noise can be found in \cite{Nyc:Szn:13}.
	
\begin{figure}[h!]
		\includegraphics[width=\textwidth]{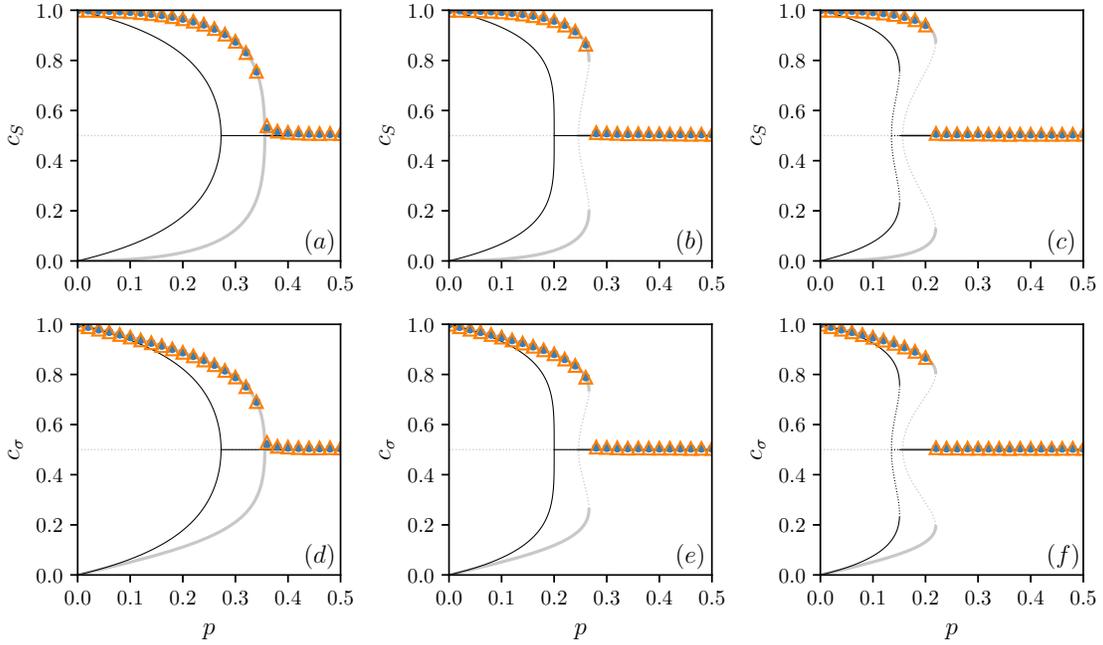}
		\caption{\textbf{Stationary concentrations of opinions as a function of independence probability $p$.} 
			Concentrations of positive public opinions are presented in upper panels (a)-(c), whereas concentrations of positive private opinions are presented in bottom panels (d)-(f). Different columns correspond to different $q$-panel sizes: (a) and (d) $q=4$, (b) and (e) $q=5$, and (c) and (f) $q=6$. Analytical results are represented by lines, solid for stable states and dotted for unstable ones: gray lines stand for AT and TA models, whereas black lines, which do not overlap symbols, stand for the original $q$-voter model with independence \cite{Nyc:Szn:Cis:12}. In general, transition points for the original $q$-voter model are located in the area of smaller values of independence than in the case of AT and TA models. Symbols represent results of Monte Carlo simulations with ordered initial conditions ($S_i(0)=\sigma_i(0)=1$ for all agents): $\triangle$ for TA (think then act) and $\bullet$ for AT (act then think) models. It is seen that TA and AT models give exactly the same outcomes. Moreover, for $q\geq 5$, the metastable region, where both order and disordered phases are stable, and they coexist, shows up. Results from Monte Carlo simulations have been obtained for the system of the size $N=10^5$ averaged over $100$ samples after $10^5$ MCS.}
		\label{fig:opinions}
	\end{figure}

	First, let us derive formulas for the time evolution of both concentrations: the public $c_S$ and the private $c_\sigma$. In the model, a random sequential updating scheme is used, so at each elementary time step only a single agent can make an action. Therefore, its behavior may lead into three situations. The agent can change its opinion, and then the concentrations may increase or decrease by $1/N$, but it can preserve its old opinion as well, and in such a case the concentrations do not change.
	Introducing the following transition rates:
	\begin{align}
	\gamma_S^+&=\text{P}\left(c_S\rightarrow c_S+\frac{1}{N}\right),\\
	\gamma_S^-&=\text{P}\left(c_S\rightarrow c_S-\frac{1}{N}\right),
	\end{align}
	for the public concentration, and the analogous formulas, $\gamma_\sigma^+$ and $\gamma_\sigma^-$, for the private one, we can write down two rate equations, which set the time evolution of the system. In the limit of $N\rightarrow\infty$, they have the following differential form
	\begin{equation}
	\label{eq:partialC}
	\frac{d c_S}{d t}=\gamma^+_S-\gamma^-_S,
	\end{equation}
	where time is measured as usual in Monte Carlo steps (MCS), and one such step corresponds to $N$ elementary time steps, so that $N$ randomly picked agents have an opportunity to reconsider their opinions. For the private opinion, the subscript in the above equation changes to $\sigma$.
	The explicit forms of transition rates are derived based on the model description and the mean-filed approximation, which neglects all fluctuations in the system. Additionally at this point, we assume that the states of public and private opinions are independent. Doing so, we ease the calculations, but on the other hand, we lose the mathematical exactness, and the transition rates for both models AT and TA become identical. Therefore, it seems that the mean-field analysis with this simplification is not able to capture the differences between our models even if they do exist. But anyway, let us proceed with the calculations and check if the results match the Monte Carlo simulations.
	Coming back to our transition rates for the public case, we have
	\begin{align}
	\label{eq:gammaS}
	\gamma^+_S&=(1-c_S)\left\{pc_\sigma+(1-p)\left[c_\sigma(1-(1-c_S)^q)+(1-c_\sigma)c_S^q\right]\right\},\\
	\gamma^-_S&=c_S\left\{p(1-c_\sigma)+(1-p)\left[c_\sigma(1-c_S)^q+(1-c_\sigma)(1-c_S^q)\right]\right\},
	\end{align}
	whereas for the private case, the formulas are as follows
	\begin{align}
	\gamma^+_\sigma&=(1-c_\sigma)\left[p/2+(1-p)c_S^q\right],\\
	\label{eq:gammaO}
	\gamma^-_\sigma&=c_\sigma\left[p/2+(1-p)(1-c_S)^q\right].
	\end{align}
	In our analysis, we are mostly interested in the final state of the system for which both concentrations have already settled at certain levels, and they do not change over time anymore. Such a state is called the stationary one. Although the stationary concentration values are time-independent, they might depend on the external parameters of the model like the level of independence $p$ or the influence group size $q$.
	These dependencies can be obtained just by solving simultaneously two equations $\gamma^+_S=\gamma^-_S$ and $\gamma^+_\sigma=\gamma^-_\sigma$ because then, the time derivatives of both concentrations (see Eq.~(\ref{eq:partialC})) vanish, check as well S1 Appendix for more detailed  mathematical guidance.
	Since the obtained formulas for the stationary values of the concentrations are rather complicated, we visualize them for 3 $q$-panel sizes in Fig.~\ref{fig:opinions}. In the same plot, we can also find corresponding results of Monte Carlo simulations indicated by symbols. 
	Surprisingly, we see that the agreement between results coming from these two approaches is significant despite our assumption about independent states of public and private opinions, and indeed, both models AT and TA produce the same final outcome, at least for the stationary values of concentrations. 
	Moreover, similarly as for the original $q$-voter model with independence, there are phase transitions between the low-independence phase in which one opinion is dominating, a positive or a negative one, at both levels (i.e., $c_S\ne 1/2$ and $c_\sigma\ne 1/2$), and the high-independence phase in which both, positive and negative, opinions are equally popular (i.e., $c_S= 1/2$ and $c_\sigma= 1/2$). 
	Note that these transitions occur at the same point on both, public and private, levels; compare plots in the same columns in Fig.~\ref{fig:opinions}. 
	In the case of continuous phase transitions, which happen for $q\leq 4$, the critical point that separates these low- and high-independence phases can be calculated as
	\begin{equation}
	p^*=\frac{1-2^{q-1}+\sqrt{(1-2^{q-1})^2+4q(q-1)}}{2q}.
	\label{eq:crit}	
	\end{equation}
	Therefore, as we stated before, for $p<p^*$ the system is ordered since one opinion wins over the other, and above the critical point, that is $p>p^*$, the system is disordered because both opinions coexist equally likely. 
	More investigative reader can find the derivation of the above formula in S1 Appendix.
	This critical point is also known for the original $q$-voter model with independence \cite{Nyc:Szn:13}. It turns out that it is shifted towards smaller values of independence in comparison to our modified version with two levels of opinions. However, this difference in critical values of independence decreases as the number of $q$-panel members increases.
	
	Furthermore, our model exhibits also discontinuous phase transitions for $q\geq 5$, so the phase transition type depends on the influence group size $q$.
	A similar behavior is displayed by the $q$-voter model with independence as well, however, in that model continuous phase transitions become discontinuous for $q=6$ \cite{Nyc:Szn:Cis:12,Nyc:Szn:13}, see also Fig.~\ref{fig:opinions} for the comparison between the models.
	
	Note that for discontinuous phase transitions, the formula for $p^*$ corresponds to the lower bound of a metastable region where the final, stationary state of the system depends on the initial conditions, and as a result, the hysteresis is observed; check the middle and the right columns of Fig~\ref{fig:opinions}.
	Looking at the same figure, one can spot that the $q$-voter model resembles in the shapes of concentrations more our modified model at the private level than at the public one. It may be surprising at first glance because we are used to think about the original $q$-voter model as a model with only a public level of opinion. However, this phenomenon can be understood if we look at the algorithms: in the case of independence a private opinion changes in exactly the same way as an opinion in the $q$-voter model, i.e., randomly, whereas a public opinion is replaced by a private one.
	
\begin{figure}[h!]
		\includegraphics[width=\textwidth]{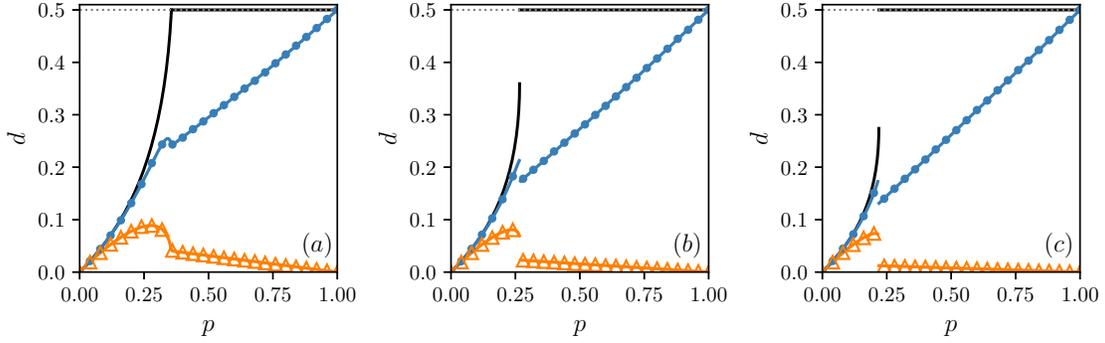}
		\caption{\textbf{Stationary value of the dissonance level as a function of independence.} Results for different $q$-panel sizes are presented: (a) $q=4$, (b) $q=5$ and (c) $q=6$. Analytical results are marked by lines: black lines, which do not overlap symbols, are obtained under the assumption that private and public opinions are independent, i.e., by Eq. (\ref{eq:d}), whereas lines that overlap symbols are obtained from the exact analytical approach. Symbols illustrate outcomes from Monte Carlo simulations: $\triangle$ for the TA (think then act) and $\bullet$ for the AT (act then think) models. The initial state of the system is the ordered one. As seen, the level of dissonance for the AT model is higher that for the TA model. Results from Monte Carlo simulations have been obtained for the system of the size $N=10^5$ averaged over $100$ samples after $10^5$ MCS.}
		\label{fig:dissonances}
	\end{figure}

	So far, it seems that the order of agents' actions does not play an important role and does not influence the final outcome of the system.
	However, when the level of dissonance in the society $d$ is measured as well, these two updating schemes produce very different results and become distinguishable on a macroscopic level, see Fig.~\ref{fig:dissonances}.
	Moreover, we see that states of private and public opinions are not completely independent as we assumed before in our calculations. If they were independent, the level of dissonance would be expressed in the simple form 
	\begin{equation}
	d=c_S(1-c_\sigma)+c_\sigma(1-c_S),
	\label{eq:d}	
	\end{equation}
	and all simulation results, from both models, would lie on the black curves that correspond to the above formula in Fig.~\ref{fig:dissonances}.
	Evidently, this is the case for small values of the independence level $p$.
	However, it seems that the results from AT model follow Eq.~(\ref{eq:d}) in a wider range of independence variability, so that the states of private and public opinions stay independent for bigger values of $p$ than in the case of the TA model.
	Moreover, we can infer that in general these public and private opinion states are more strongly correlated in TA model since the simulated points related to this updating order lie further away from Eq.~(\ref{eq:d}) than the points from the AT model.
	
\begin{figure}[h!]
		\includegraphics[width=\textwidth]{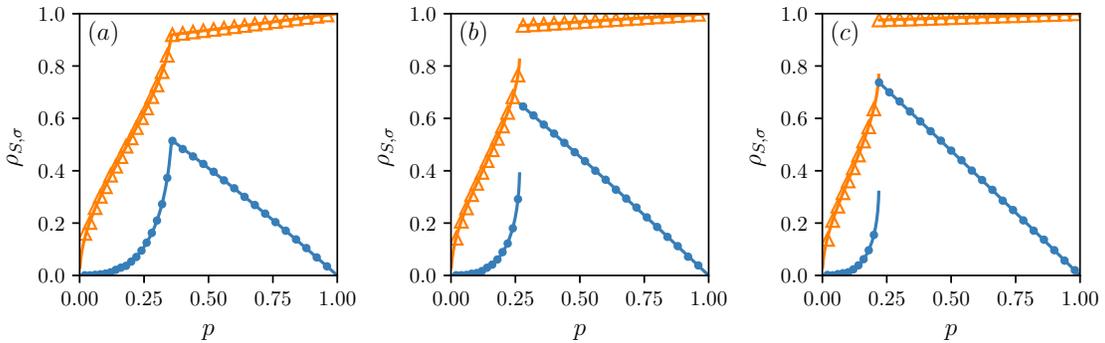}
		\caption{\textbf{Pearson correlation coefficient between public and private opinions.} Results for different $q$-panel sizes are presented: (a) $q=4$, (b) $q=5$ and (c) $q=6$. Analytical results are marked by lines. Symbols illustrate outcomes from Monte Carlo simulations: $\triangle$ for the TA (think then act) and $\bullet$ for the AT (act then think) models. The system starts its evolution from the ordered state for all cases. We can see that the states of opinions on public and private levels are more strongly correlated in the TA model. On the other hand, for small values of the independence level, the correlation coefficient stays close to zero in a wider range of independence variability for the AT model. Note that in the high-independence phase, our measure of association increases to 1 for the TA model whereas it decreases to 0 in the case of the AT model. Results from Monte Carlo simulations have been obtained for the system of the size $N=10^6$ after $10^3$ MCS. }
		\label{fig:corr}
	\end{figure}

	Since we already know that states of our opinions on private and public levels are coupled and are not fully independent of each other, we can repeat our mean-field analysis; however, this time without our false assumption. 
	This will make our calculations strict for an infinite complete graph, and as a result, will allow us to obtain accurate predictions about the dissonance levels in both models.
	In order to capture the fact that states of private and public opinions may depend on each other, we have to describe the system more rigorously. Therefore, we split our two concentrations, $c_S$ and $c_\sigma$, into four: $c_{\uparrow\uparrow}$, $c_{\uparrow\downarrow}$, $c_{\downarrow\downarrow}$, and $c_{\downarrow\uparrow}$  -- one for each state combination of public and private opinions, where the first subscript refers to the public opinion, and the second to the private one. Two arrow directions, $\uparrow$ and $\downarrow$, correspond to two different opinion states, $1$ and $-1$.
	In such a notation we have
	\begin{align}
	c_S&=c_{\uparrow\uparrow}+c_{\uparrow\downarrow},\label{eq:newNotation1}\\
	c_\sigma&=c_{\uparrow\uparrow}+c_{\downarrow\uparrow},\label{eq:newNotation2}\\
	d&=c_{\uparrow\downarrow}+c_{\downarrow\uparrow},
	\label{eq:newNotation3}
	\end{align}
	note, however, that this is only another way of representing $c_S$, $c_\sigma$, and $d$, and the initial definitions of these quantities Eqs.~(\ref{eq:def:pub})-(\ref{eq:def:dis}) are still valid.
	Additionally, we know that $c_{\uparrow\uparrow}+c_{\uparrow\downarrow}+c_{\downarrow\downarrow}+c_{\downarrow\uparrow}=1$ since these four combinations of opinions exhaust all the possibilities.
	Further steps are analogical to the previous ones -- we write down transition rates for all concentrations separately and solve the system of equations for the stationary states. 
	However, this time, formulas for the transition rates are more complicated, so we omit their presentation herein.
	All readers interested in the precise forms of them are referred to S2 Appendix.
	The results of the above, exact, analytical approach are also presented in Fig. \ref{fig:dissonances} along with the previous outcomes from the simplified method and Monte Carlo simulations.
	Now, our calculations correctly predict the behavior of the dissonance levels for both models in the whole domain of the independence values $p$.
	This approach also produces the same stationary values of the concentrations as the simplified method, which does not surprise since we have known that the simulated stationary concentrations were already in agreement with our previous calculations.
	Additionally, at this stage, we can say a bit more about correlations between public and private opinion states and compute the Pearson correlation coefficient $\rho_{S,\sigma}$ for these variables analytically:
	\begin{equation}
	\rho_{S,\sigma}=\frac{c_S(1-c_\sigma)+c_\sigma(1-c_S)-d}{2\sqrt{c_Sc_\sigma(1-c_S)(1-c_\sigma)}},
	\end{equation}
	a full derivation is presented in S3 Appendix. Based on the above formula, two remarks can be made right away. The first is that the correlations between public and private opinions should be smaller for AT model. This is because the level of dissonance $d$, which appears with the minus sign in the above equation, is higher for this updating order, and at the same time the stationary concentrations in both models are the same. Furthermore, if the states of public and private opinions are independent then formula (\ref{eq:d}) is correct, which leads to $\rho_{S,\sigma}=0$. Figure~\ref{fig:corr} illustrates the dependency between this correlation coefficient calculated in the stationary state and the level of independence $p$ for the systems composed of $N=10^6$ agents with different sizes of the $q$-panel. Indeed, the states of public and private opinions are stronger associated with each other in the TA model. Moreover, we see that the correlation coefficient is a non-decreasing function of $p$ only in the case of TA model.
	
	It is not so difficult to understand why the correlations between public and private opinions are higher and simultaneously dissonance is lower for TA than for AT model if we look once again at algorithms. This situation is due to independence that acts differently on opinions on public and private levels. Within AT model, the private opinion is updated last, and then it can randomly change its state with probability $p$. Hence, independence may destroy the agreement between private and public opinions at the end of a time step increasing the dissonance. It is easy to see that for the extreme case when $p=1$. Then in the last step independence occurs, and the private opinion takes value 1 or -1 with equal chances according to the algorithm. However, these are also the only acceptable states for the public opinion, hence, with probability 1/2 our opinions anti-align resulting in the dissonance value $d=1/2$. On the other hand, within TA model, it is the public opinion that is updated last, and then with probability $p$ the value of the private opinion is copied to the public one. Therefore, in this case independence helps to align the states of both opinions lowering the dissonance. Again it is easy to see that for $p=1$ since then at the end of each time step the value of the private opinion is passed to the public one. Hence, both opinion are always in agreement, and there is no dissonance in the system $d=0$.
	\\
	More intriguing is the fact that in spite of significant correlations between public and private opinions, the simple mean-field approach, which assumes independence between both opinions, thus, neglects these correlations, gives correct values of the stationary concentrations. Additionally, these concentrations are identical for TA and AT models, which is also confirmed by Monte Carlo simulations. Moreover, as we have already written, our second method, which takes into account these correlations, leads exactly to the same result as the naive MFA. The question is why the model at the level of stationary opinions reduces to the simple mean-field-like model. Unfortunately, we do not have any straightforward answer to this question. Our only intuition is that this result is not universal and may be valid solely on a complete graph. If the intuition is correct, we should observe differences in the stationary concentrations between AT and TA models on other graphs, and therefore, the model examination on alternative networks is the task for the future.
	
\begin{figure}[h!]
		\includegraphics[width=\textwidth]{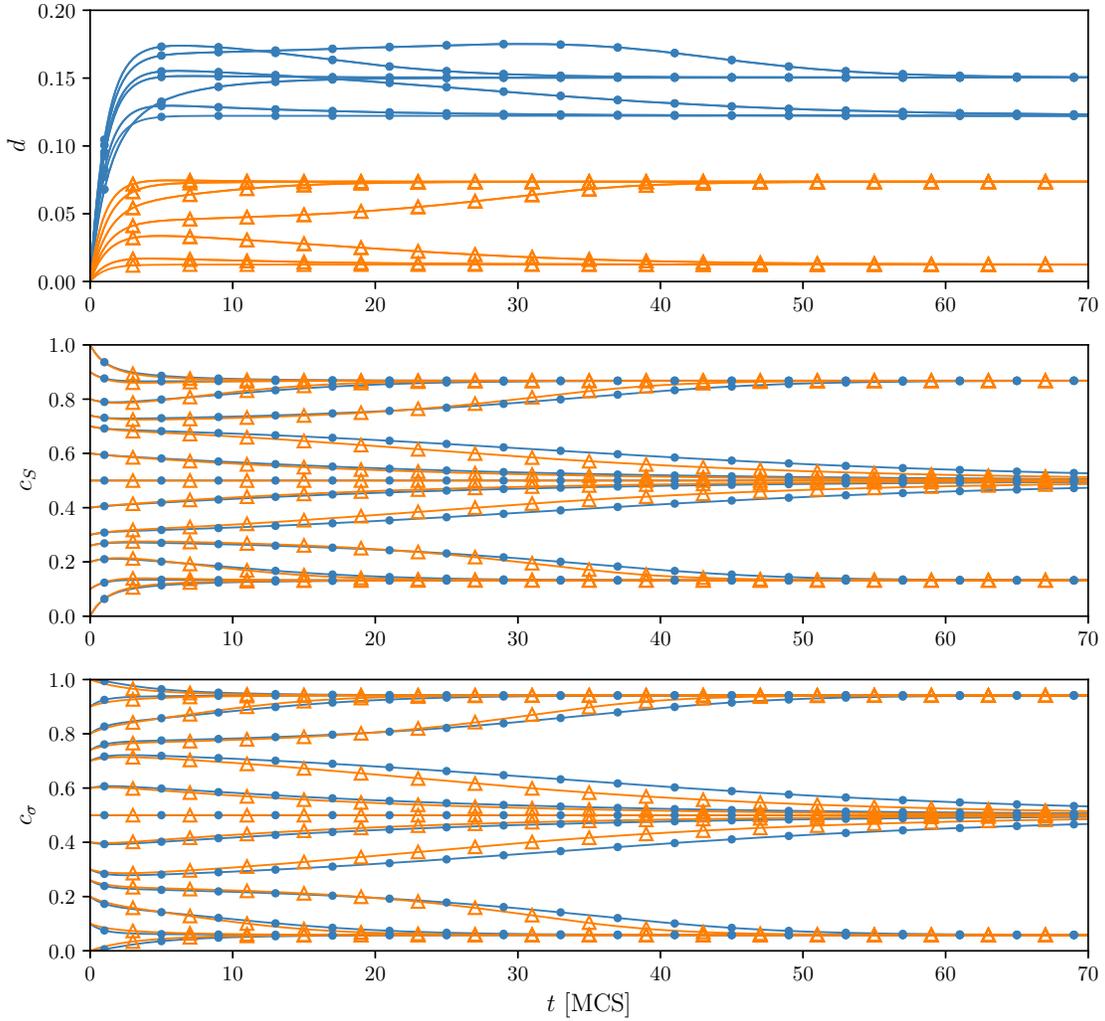}
		\caption{\textbf{Average trajectories of the dissonance and concentration of positive opinions. } Results for the dissonance are presented in top panels, whereas results for public and private opinions in the middle and bottom panels respectively. Analytical results are marked by solid lines. Symbols illustrate outcomes from Monte Carlo simulations: $\triangle$ for the TA (think then act) and $\bullet$ for the AT (act then think) models.  The system is comprised of $N=10^6$ agents with the independence level $p=0.2$ and $q=6$ in all cases. Note that these parameters correspond to the metastable region where order ($c\ne 1/2$) and disorder phases ($c= 1/2$) coexist, so that the hysteresis appears, see Fig.~\ref{fig:opinions}. These two different phases and the dependence between initial and final states of the system is clearly seen in the figures. The trajectories are averaged over 100 samples.}
		\label{fig:tra}
	\end{figure}

\begin{figure}[h!]
		\includegraphics[width=\textwidth]{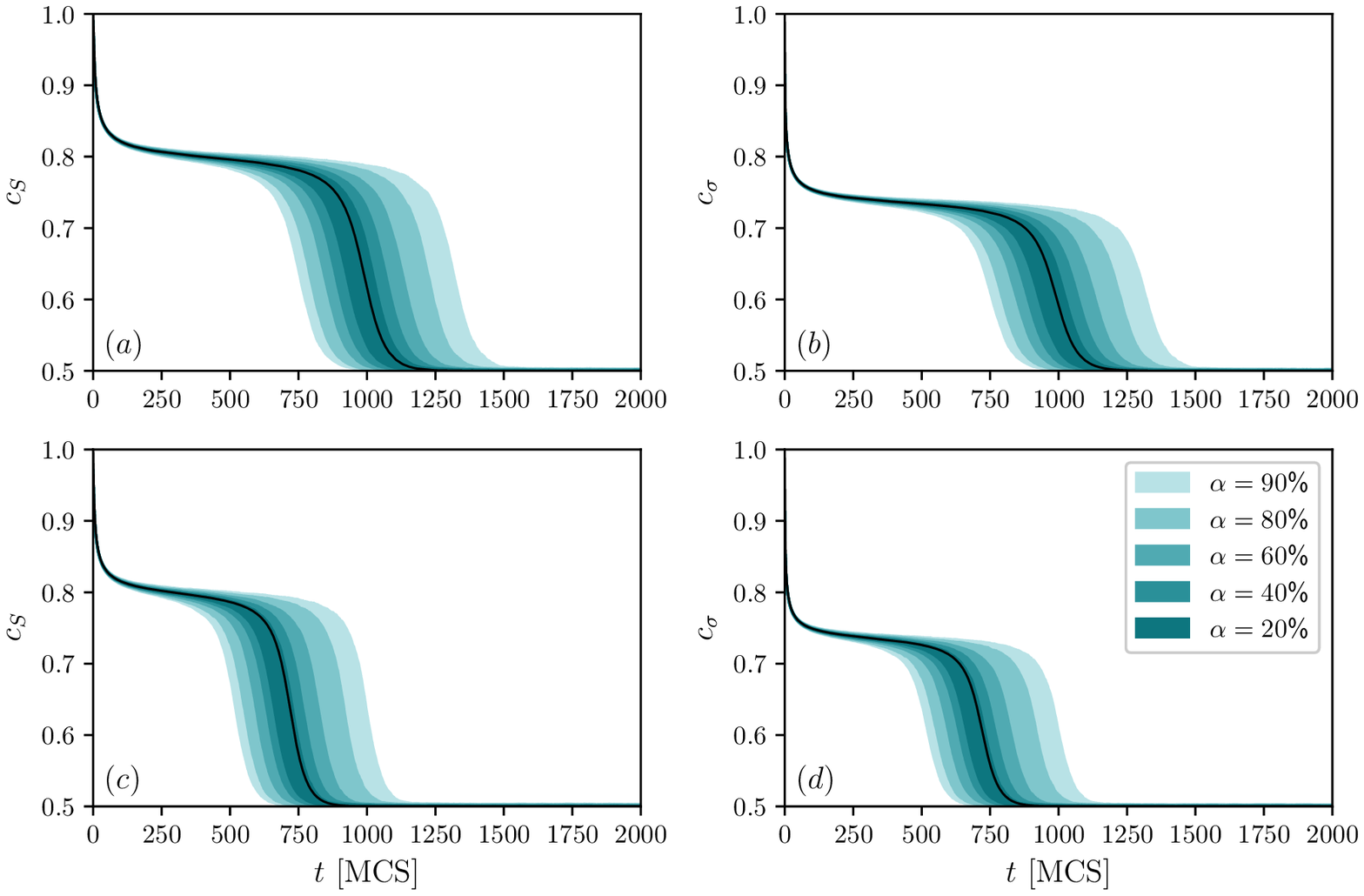}
		\caption{\textbf{Empirical quantiles of sample trajectories of public and private opinions.}
			Results for public opinions are presented on the left panels (a) and (c), whereas results for private opinions on the right panels (b) and (d). Top panels (a)-(b) correspond to TA and bottom panels (c)-(d) to the AT updating schemes. Marked with colors are the intervals containing $\alpha$ realizations closest to the median (for example, when $\alpha=80\%$ the lower bound of the marked area is the 10-th percentile, and the upper bound is the 90-th percentile). The quantiles were calculated for each time point separately. Black curves are the average trajectories based on an analytical description of the system. The simulation results were obtained using $10^3$ independent trajectories for the system of $N=10^6$ nodes, parameter $q=5$ and $p=0.2664$. The initial state of the system is $S_i(0)=\sigma_i(0)=1$ for all agents.}
		\label{fig:trajectories}
	\end{figure}

	In addition to the stationary concentrations, we can derive the whole time evolution of $c_S$,  $c_{\sigma}$, and $d$ just by solving the rate equations within the new approach. 
	Of course, such an evolution can be also easily obtained from Monte Carlo simulations.
	In Fig.~\ref{fig:tra}, we present several trajectories of the system comprised of $N=10^6$ agents averaged over 100 samples. The parameters were chosen in such a way that the considered model is in a metastable region where the dependence between initial and final states of the system is observed. 
	Note that time trajectories of the concentrations are only slightly different for TA and TA models, and eventually they reach the same final state independently of the model. Thus, only the dissonance levels differentiate between these two updating orders in the stationary state.
	For the chosen set of parameters, the analytical paths accord significantly with the average sample trajectories. 
	However, one should remember that our calculation were done in the limit of an infinite network size, so for smaller system sizes more discrepancies may originate. 
	Small systems also induce larger fluctuations in the concentrations and the dissonance level, thus, obtaining so accurate and smooth averaged trajectories may require bigger samples.
	Moreover, these fluctuations present only in finite systems cause sometimes spontaneous transitions between stationary states that might also produce some divergence in results.
	In order to show how trajectories from the same initial conditions may vary, we calculate empirical percentiles for the system of the size $N=10^6$ based on $10^3$ simulated paths, the outcomes are presented in Fig.~\ref{fig:trajectories}. 
	As seen, the variability of the trajectories might be substantial, and it is seemingly bigger for the AT model, at least for this set of parameters.
	Moreover, for this setup, both models support a long period of time without substantial changes in the concentrations. This can lead to the false conclusion that the system has already reached its final state despite the fact that the big turnover has yet to appear. 
	
	\section*{Discussion}
	We have proposed a new model of opinion dynamics  that combines (a) ideas introduced within the $q$-voter model with noise, which is an ABM and belongs to the world of sociophysics \cite{Nyc:Szn:Cis:12},  with (b) the descriptive, four-dimensional model of social response formulated by social psychologists \cite{Nai:Mac:Lev:00}. 
	In our model, agents' opinions are described by binary, dynamical variables that can be measured at two different levels,  a public and a private one.
	The source of influence is formed by the unanimous group of $q$ individuals, like in the original $q$-voter model \cite{Cas:Mun:Pas:09,Nyc:Szn:Cis:12}. 
	Not having found a clear answer to the question ``What comes first -- acting or thinking?'' in social experiments, we considered two versions of the same model that differ only by the updating sequence of opinions just to check whether this order is important from the point of view of agent-based modeling.
	In our analysis, two independent methods have been used: Monte Carlo simulations and analytical calculations within the mean-field approximation.
	Having at least two autonomous approaches is especially important from the perspective of rigorous agent-based modeling since it allows for the precise verification of the implemented model \cite{Ran:Rus:11}.
	The results obtained from both of our methods support each other well. This is not very surprising because we have considered the model on a complete graph, which corresponds to the mean field approach. Agreement between simulation results and analytical approach is in our opinion the best method of verification if the model was implemented correctly. Therefore, the complete graph should be highly recommended as a first step in agent-based modeling. Nevertheless, we are aware that investigating the model on different, complex networks will be a desired task for the future, as discussed below.

	It has been shown that the stationary values of the concentrations on the public and private levels are the same for both updating orders, i.e., for TA and AT models. Thus, both models exhibit the same order-disorder phase transitions, compare plots in Fig.~\ref{fig:opinions}.
	Their type depends on the number of individuals that exert social pressure, namely, on the $q$-panel size.
	For $q\leq4$, phase transitions are continuous, and we know the analytical formula for the critical point that separates low- and high-independence phases, see Eq.~(\ref{eq:crit}). 
	On the other hand, for $q\geq 5$, phase transitions become discontinuous, and a metastable region appears where ordered and disordered phases coexist, and final states of the system depend on initial conditions. In this case, Eq.~(\ref{eq:crit}) corresponds to the lower bound of this metastable region.
	A similar change of a phase transition type is also exhibited by the original $q$-voter model with independence, which incorporates only one opinion \cite{Nyc:Szn:Cis:12}. However, in that case, continuous phase transitions become discontinuous for $q=6$.
	So, it seems like adding to the original model another opinion level coupled with the old one lowers the point of the transition type change. Interestingly, introducing another layer of a network drives this point down as well when we consider the $q$-voter model on multiplex networks \cite{Chm:Szn:15}.
	
	Since AT and TA updating schemes give the same stationary concentrations, we are not able to differentiate between models on a macroscopic scale only by measuring a fraction of individuals with a positive opinion on one or the other level in the final state. 
	This may lead to the false conclusion that the updating order does not play an important role, and although the models differ on a microscopic scale, these differences are so subtle that the same collective behavior is produced in both cases.
	However, the dissonance level is significantly different for both updating schemes.
	For instance, the final level of dissonance is higher in the case of AT model. Nevertheless, the differences are not only quantitative. In the high-independence phase, the dissonance increases with the independence probability $p$ for the AT model whereas it decreases to zero in the case of the TA model, look at Fig.~\ref{fig:dissonances}. Finally, we found out that the states of public and private opinions are in general less correlated in the AT model, and for small values of $p$, they stay independent in a wider range of this parameter variability, analyze Figs.~\ref{fig:dissonances} and \ref{fig:corr}.
	
	Based on our study, we can state that, indeed, the updating order of public and private opinions matters and changes the general behavior of the system. However, some model features remain unchanged, like the stationary values of the concentrations.
	From the perspective of our model, thinking then acting is more profitable in the sense that it leads to smaller dissonance in the society, and it makes public and private opinions more strongly correlated. One could also conclude that our results indicate that TA model is more correct than AT because it seems logical that in the case of high independence, i.e., low social pressure, people should behave in a way they think so the dissonance should be low and correlation between public and private opinion high, which is exactly the case of TA model.

	Nevertheless, these are the results for a specific, relatively simple, model with several assumptions that not always have to reflect the reality.
	Therefore, taking into account other factors may lead to different conclusions.
	Luckily, agent-based modeling allows us to tune the conditions of simulations and make the system more realistic by changing agent attributes and rules of behaviors, introducing different system space or adding external environment influence \cite{Jun:etal:18}.
	
	So, let us point out a few such assumptions that could be investigated in future research. 
	First of all, we have done our analysis on complete graphs.
	Although these structures mimic well relationships in small communities where all members know each other, they are inappropriate to model large societies, which are said to belong to the class of co called complex networks \cite{Boc:etal:06}. Additional analysis on such structures could reveal more differences between the updating schemes. However, obtaining analytical results in that case might require more sophisticated methods that are specially dedicated to network problems like heterogeneous mean-field or pair approximations \cite{Mor:etal:13, Pug:Cas:09, Jed:17}. The original $q$-voter model with independence has been already studied on complex structure \cite{Jed:17}, and it turned out that the average node degree plays an important role in the model behavior.
	Another issue has been already mentioned in the model description and concerns the way we decide about the behavior of agents. Herein, we considered only the situation-oriented approach in which each agent could be either conformist or independent at different times. The person-oriented approach, where the agents’ way of acting does not change across time and settings, has not been investigated as yet. Still, it could drastically modify the model behavior, just as it did in the case of the $q$-voter model  \cite{Jed:Szn:17}.
	
	Finally, we assumed that the size of the influence group $q$ is the same for all agents and does not change during one simulation. This makes our population homogeneous whereas in the reality $q$ may vary \cite{Rad:etal:17,Mel:Mob:Zia:17}, thus, it would be interesting to introduce more heterogeneous agents to the system. Another more realistic modification of the model, also connected with the influence group, is to resign from requiring the unanimity of the $q$-panel. Instead, one could allow agents to follow the group when a minimal number of its members with the same opinion is crossed, like in the threshold $q$-voter model \cite{Nyc:Szn:13,Vie:Ant:18}. 
	
	From all the above, we see that there are still many important additional factors that are particularly relevant to social agent-based modeling. These factors should be considered in future analyses.
	
	\section*{Acknowledgments}
	This work was partially supported by the National Science Center (NCN, Poland) through grants no. 2016/21/B/HS6/01256 and no. 2016/23/N/ST2/00729.
	
	
	%
	%
	%
	
	\bibliography{literature}
	
	\section*{Supporting information}
	\textbf{S1 Appendix. Derivation of formulas for the critical points and stationary states.}\newline\newline
	\textbf{S2 Appendix. Transition rates in the exact mathematical formulations of the models.}\newline\newline
	\textbf{S3 Appendix. Pearson correlation coefficient derivation.}
	
	\newpage
	\section*{Supporting information: S1 Appendix.} 
	\vspace{-0.25cm}
	\textbf{Derivation of formulas for the critical points and stationary states.}\\~\\
	In the main text, we include an expression for the critical point $p^*$ that separates low- and high-independence phases in the case of continuous phase transitions and becomes the lower bound of a metastable region in the case of discontinuous phase transitions.
	Here, we show how to derive the formula for this specific point. Presented method is quite general and can be applied to other similar models.
	First, let us recall the transition rates for the concentrations of positive public opinions
	\begin{align}
	\gamma^+_S&=(1-c_S)\left\{pc_\sigma+(1-p)\left[c_\sigma(1-(1-c_S)^q)+(1-c_\sigma)c_S^q\right]\right\},\\
	\gamma^-_S&=c_S\left\{p(1-c_\sigma)+(1-p)\left[c_\sigma(1-c_S)^q+(1-c_\sigma)(1-c_S^q)\right]\right\},
	\end{align}
	and private ones
	\begin{align}
	\gamma^+_\sigma&=(1-c_\sigma)\left[p/2+(1-p)c_S^q\right],\label{eq:gammaO1}\\
	\gamma^-_\sigma&=c_\sigma\left[p/2+(1-p)(1-c_S)^q\right].\label{eq:gammaO2}
	\end{align}
	Now, for each opinion level, let us define a quantity that can be thought of as net force acting on our system, which can increase or decrease corresponding concentration of positive opinions:
	\begin{align}
	F_S(c_S,c_\sigma,p)=\gamma^+_S-\gamma^-_S,\\
	F_\sigma(c_S,c_\sigma,p)=\gamma^+_\sigma-\gamma^-_\sigma.
	\end{align}
	We treat $q$ as a fixed parameter, so we do not include it in arguments of the above functions.
	Naturally, in the stationary state, our forces are equal to zero $F_S(c_S,c_\sigma,p)=0$ and $F_\sigma(c_S,c_\sigma,p)=0$. This creates two conditions for the stationary values of $c_S$ and $c_\sigma$. 
	Moreover, these stationary values of concentrations are coupled, and we can easily derive this dependency using the above vanishing force condition for the private opinion level together with Eqs.~(\ref{eq:gammaO1}) and (\ref{eq:gammaO2}). This procedure leads to the expression for the stationary values of $c_\sigma$ in the following form
	\begin{equation}
	c_\sigma=\frac{\frac{1}{2}p+(1-p)c_S^q}{p+(1-p)\left[c_S^q+(1-c_S)^q\right]}.
	\label{eq:csigma}
	\end{equation}
	As we see, in the stationary state, $c_\sigma(c_S,p)$ depends only on $c_S$ and $p$. Therefore, in this state, the force related to the public opinion level is, in fact, a two-variable function that equals to zero 
	\begin{equation}
	F_S(c_S,p)=0
	\label{eq:Fs}
	\end{equation}
	since we can eliminate $c_\sigma$ from the formulas using Eq.~(\ref{eq:csigma}).
	Note that the above expression is also an implicit equation for the stationary values of the concentration of positive opinions at the public level $c_S$ where
	\begin{equation}
	F_S(c_S,p)=\frac{1}{2}\left[(1-c_S)^q-c_S^q\right]p^2-\frac{1}{2}(1-c_S^q-(1-c_S)^q)(2c_S-1)p+(1-c_S)c_S^q-c_S(1-c_S)^q.
	\end{equation}
	Now, differentiating Eq.~(\ref{eq:Fs}) with respect to the concentration gives
	\begin{equation}
	\frac{\partial F_S(c_S,p)}{\partial c_S}+\frac{\partial F_S(c_S,p)}{\partial p}\frac{dp}{dc_S}=0.
	\label{eq:differentialEq}
	\end{equation}
	But since the critical point corresponds to the extremum of $p(c_S)$, the first derivative of the independence level over $c_S$ vanishes at $(c_S,p)=(1/2,p^*)$, so that we have
	\begin{equation}
	\left.\frac{dp}{dc_S}\right|_{(1/2,p^*)}=0.
	\end{equation}
	Hence, evaluating Eq.~(\ref{eq:differentialEq}) at the critical point gives the condition for its derivation
	\begin{equation}
	\left.\frac{\partial F_S(c_S,p)}{\partial c_S}\right|_{(1/2,p^*)}=0.
	\end{equation}
	Following the above instructions, one can arrive at the quadratic equation for the critical point
	\begin{equation}
	q(p^*)^2+(2^{q-1}-1)p^*-q+1=0.
	\end{equation}
	Finally, taking into account only a positive root, since $p^*>0$, results in the following formula
	\begin{equation}
	p^*=\frac{1-2^{q-1}+\sqrt{(1-2^{q-1})^2+4q(q-1)}}{2q}
	\end{equation}
	presented in the main text.
	
	\newpage
	\section*{Supporting information: S2 Appendix.} 
	\vspace{-0.25cm}
	\textbf{Transition rates in the exact mathematical formulations of the models.}\\~\\
	Herein, we present expressions for all transition rates related to four concentrations $c_{\uparrow\uparrow}$, $c_{\uparrow\downarrow}$, $c_{\downarrow\downarrow}$, and $c_{\downarrow\uparrow}$ for both updating scheme orders of private and public opinions. These four quantities fully describe the models' behavior since based on them we can obtain the concentrations of positive opinions on public and private levels as well as the dissonance level using the following formulas mentioned already in the main text:
	\begin{align}
	c_S&=c_{\uparrow\uparrow}+c_{\uparrow\downarrow},\\
	c_\sigma&=c_{\uparrow\uparrow}+c_{\downarrow\uparrow},\\
	d&=c_{\uparrow\downarrow}+c_{\downarrow\uparrow}.
	\end{align}
	We recall that in our notation convention the first lower index, that is to say, the first arrow, corresponds to the public opinion state whereas the second one to the private opinion state. The arrow direction, $\uparrow$ or $\downarrow$, corresponds to two different opinion states, 1 or −1.
	\\~\\
	\noindent \underline{Think then act (TA) model:}
	\begin{align}
	\gamma_{\uparrow\uparrow}^+=&(c_{\uparrow\downarrow}+c_{\downarrow\downarrow})\left[\frac{p^2}{2}+\frac{p}{2}(1-p)(1-(1-c_S)^q)+p(1-p)c_S^q+(1-p)^2c_S^q(1-(1-c_S)^q)\right]\nonumber\\
	&+c_{\downarrow\uparrow}\left[\frac{p^2}{2}+\frac{3p}{2}(1-p)(1-(1-c_S)^q)+(1-p)^2(1-(1-c_S)^q)^2\vphantom{\frac{p}{2}}\right],\\
	\gamma_{\uparrow\uparrow}^-=&c_{\uparrow\uparrow}\left[\frac{p}{2}(1-p)(1-c_S)^q+\frac{p}{2}+(1-p)(1-c_S)^q+(1-p)^2(1-(1-c_S)^q)(1-c_S)^q\vphantom{\frac{p}{2}}\right],\\
	\gamma_{\uparrow\downarrow}^+=&(c_{\uparrow\uparrow}+c_{\downarrow\uparrow})\left[\frac{p}{2}(1-p)c_S^q+(1-p)^2(1-c_S)^qc_S^q\right]+c_{\downarrow\downarrow}\left[\frac{p}{2}(1-p)c_S^q+(1-p)^2(1-c_S^q)c_S^q\right],\\
	\gamma_{\uparrow\downarrow}^-=&c_{\uparrow\downarrow}\left[\frac{p^2}{2}+\frac{3p}{2}(1-p)(1-c_S^q)+\frac{p}{2}+(1-p)c_S^q+(1-p)^2(1-c_S^q)^2\vphantom{\frac{p}{2}}\right],\\
	\gamma_{\downarrow\downarrow}^+=&(c_{\uparrow\uparrow}+c_{\downarrow\uparrow})\left[\frac{p^2}{2}+\frac{p}{2}(1-p)(1-c_S^q)+p(1-p)(1-c_S)^q+(1-p)^2(1-c_S)^q(1-c_S^q)\right]\nonumber\\
	&+c_{\uparrow\downarrow}\left[\frac{p^2}{2}+\frac{3p}{2}(1-p)(1-c_S^q)+(1-p)^2(1-c_S^q)^2)\vphantom{\frac{p}{2}}\right],\\
	\gamma_{\downarrow\downarrow}^-=&c_{\downarrow\downarrow}\left[\frac{p}{2}(1-p)c_S^q+\frac{p}{2}+(1-p)c_S^q+(1-p)^2(1-c_S^q)c_S^q\vphantom{\frac{p}{2}}\right],\\
	\gamma_{\downarrow\uparrow}^+=&c_{\uparrow\uparrow}\left[\frac{p}{2}(1-p)(1-c_S)^q+(1-p)^2(1-(1-c_S)^q)(1-c_S)^q\right]\nonumber\\
	&+(c_{\uparrow\downarrow}+c_{\downarrow\downarrow})\left[\frac{p}{2}(1-p)(1-c_S)^q+(1-p)^2c_S^q(1-c_S)^q\right],\\
	\gamma_{\downarrow\uparrow}^-=&c_{\downarrow\uparrow}\left[\frac{p^2}{2}+\frac{3p}{2}(1-p)(1-(1-c_S)^q)+\frac{p}{2}+(1-p)(1-c_S)^q+(1-p)^2(1-(1-c_S)^q)^2\right].
	\end{align}
	\underline{Act then think (AT) model:}
	\begin{align}
	\gamma_{\uparrow\uparrow}^+=&(c_{\uparrow\downarrow}+c_{\downarrow\downarrow})\left[\frac{p}{2}(1-p)c_S^q+(1-p)^2c_S^{2q}\right]+c_{\downarrow\uparrow}\left[\frac{p^2}{2}+\frac{3p}{2}(1-p)(1-(1-c_S)^q)\right.\nonumber\\
	&\left.+(1-p)^2(1-(1-c_S)^q)^2\right],\\
	\gamma_{\uparrow\uparrow}^-=&c_{\uparrow\uparrow}\left[\frac{p^2}{2}+p(1-p)(1-c_S)^q+(1-p)(1-c_S)^q+\frac{p}{2}(1-p)(1-(1-c_S)^q)\right.\nonumber\\
	&\left.+(1-p)^2(1-(1-c_S)^q)(1-c_S)^q\vphantom{\frac{p}{2}}\right],\\
	\gamma_{\uparrow\downarrow}^+=&(c_{\uparrow\uparrow}+c_{\downarrow\uparrow})\left[\frac{p^2}{2}+p(1-p)(1-c_S)^q+\frac{p}{2}(1-p)(1-(1-c_S)^q)\right.\nonumber\\
	&\left.+(1-p)^2(1-(1-c_S)^q)(1-c_S)^q\vphantom{\frac{p}{2}}\right]+c_{\downarrow\downarrow}\left[\frac{p}{2}(1-p)c_S^q+(1-p)^2c_S^q(1-c_S^q)\right],\\
	\gamma_{\uparrow\downarrow}^-=&c_{\uparrow\downarrow}\left[p+(1-p)(1-c_S^q)+\frac{p}{2}(1-p)c_S^q+(1-p)^2c_S^{2q}\right],\\
	\gamma_{\downarrow\downarrow}^+=&(c_{\uparrow\uparrow}+c_{\downarrow\uparrow})\left[\frac{p}{2}(1-p)(1-c_S)^q+(1-p)^2(1-c_S)^{2q}\right]+
	c_{\uparrow\downarrow}\left[\frac{p^2}{2}+\frac{3p}{2}(1-p)(1-c_S^q)\right.\nonumber\\
	&\left.+(1-p)^2(1-c_S^q)^2\vphantom{\frac{p}{2}}\right],\\
	\gamma_{\downarrow\downarrow}^-=&c_{\downarrow\downarrow}\left[\frac{p^2}{2}+p(1-p)c_S^q+(1-p)c_S^q+\frac{p}{2}(1-p)(1-c_S^q)+(1-p)^2(1-c_S^q)c_S^q\right],\\
	\gamma_{\downarrow\uparrow}^+=&c_{\uparrow\uparrow}\left[\frac{p}{2}(1-p)(1-c_S)^q+(1-p)^2(1-c_S)^q(1-(1-c_S)^q)\right]\nonumber\\
	&+(c_{\uparrow\downarrow}+c_{\downarrow\downarrow})\left[\frac{p^2}{2}+p(1-p)c_S^q+\frac{p}{2}(1-p)(1-c_S^q)+(1-p)^2(1-c_S^q)c_S^q\right],\\
	\gamma_{\downarrow\uparrow}^-=&c_{\downarrow\uparrow}\left[p+(1-p)(1-(1-c_S)^q)+\frac{p}{2}(1-p)(1-c_S)^q+(1-p)^2(1-c_S)^{2q}\right].
	\end{align}
	
	\newpage
	\section*{Supporting information: S3 Appendix.} 
	\vspace{-0.25cm}
	\textbf{Pearson correlation coefficient derivation.}\\~\\
	Our starting point for deriving the explicit formula for the Pearson correlation coefficient $\rho_{S,\sigma}$ between states of public $S$ and private $\sigma$ opinions is its definition:
	\begin{equation}
	\rho_{S,\sigma}=\frac{\textrm{Cov}[S,\sigma]}{\sqrt{\textrm{Var}[S]\textrm{Var}[\sigma]}}.
	\label{eq:rho}
	\end{equation}
	After applying the formulas for the covariance
	\begin{equation}
	\textrm{Cov}[S,\sigma]=\textrm{E}[S\sigma]-\textrm{E}[S]\textrm{E}[\sigma],
	\label{eq:cov}
	\end{equation}
	and the variances of both opinions
	\begin{align}
	\textrm{Var}[S]=&\textrm{E}[S^2]-\textrm{E}[S]^2, \label{eq:VarS}\\
	\textrm{Var}[\sigma]=&\textrm{E}[\sigma^2]-\textrm{E}[\sigma]^2,
	\end{align}
	the calculations boil down to finding all the above expected values.
	Both opinions can take one of two possible states 1 or -1. The probabilities of these events are equal to $\textrm{P}(S=1)=c_S$ and $\textrm{P}(S=-1)=1-c_S$, with analogical formulas for the private opinion $\sigma$.
	Having the outcomes of our opinions and their occurrence probabilities, we can easily calculate their expected values, for the public opinion we have:
	\begin{align}
	\textrm{E}[S]&=\textrm{P}(S=1)-\textrm{P}(S=-1)=2c_S-1,\\
	\textrm{E}[S^2]&=\textrm{P}(S=1)+\textrm{P}(S=-1)=1.
	\end{align}
	Hence, after taking into account Eq.~(\ref{eq:VarS}), the variance of $S$ has the following form
	\begin{equation}
	\textrm{Var}[S]=1-(2c_S-1)^2=4c_S(1-c_S).
	\end{equation}
	Now, the last missing part is the expected value of the product of our two opinions $\textrm{E}[S\sigma]$, but since $S$ and $\sigma$ are not independent, we have to use our four concentrations $c_{\uparrow\uparrow}$, $c_{\uparrow\downarrow}$, $c_{\downarrow\downarrow}$, and $c_{\downarrow\uparrow}$ that take into account the joint state of public and private opinions.
	Then we can write that $\textrm{P}(S=1,\sigma=1)=c_{\uparrow\uparrow}$, and so on for the other state combinations.
	Having this, one can arrive at the following formula for the expected value of the product of public and private opinions:
	\begin{equation}
	\textrm{E}[S\sigma]=c_{\uparrow\uparrow}-c_{\uparrow\downarrow}+c_{\downarrow\downarrow}-c_{\downarrow\uparrow}.
	\label{eq:ESsigma}
	\end{equation}
	Using the fact that all the four concentrations must sum up to one $c_{\uparrow\uparrow}+c_{\uparrow\downarrow}+c_{\downarrow\downarrow}+c_{\downarrow\uparrow}=1$ and the formula for the dissonance $d=c_{\uparrow\downarrow}+c_{\downarrow\uparrow}$, we can express Eq.~(\ref{eq:ESsigma}) in terms of the dissonance alone:
	\begin{equation}
	\textrm{E}[S\sigma]=1-2d.
	\end{equation}
	Finally, joining all the above results and using Eq.~(\ref{eq:rho}), we get the formula for the correlation coefficient presented in the main text
	\begin{equation}
	\rho_{S,\sigma}=\frac{c_S(1-c_\sigma)+c_\sigma(1-c_S)-d}{2\sqrt{c_Sc_\sigma(1-c_S)(1-c_\sigma)}}.
	\end{equation}
	
\end{document}